\documentclass[12pt]{article}
\usepackage{graphicx}
\usepackage{epsfig}
\usepackage{cite}

\begin{document}

\author{Ramon Reigada \\
Facultat de Qu\'imica, Universitat de Barcelona, \\ 
C/ Mart\'i i Franqu\`es 1, 08028 Barcelona, Spain \\
e-mail: reigada@ub.edu
\and 
Igor M. Sokolov \\
Institut f\"ur Physik, Humboldt-Universit\"at zu Berlin, \\
Newtonstr. 15, D-12489 Berlin, Germany\\
e-mail: igor.sokolov@physik.hu-berlin.de}
\title{Steady-State Luminescence of Polymers: Effects of Flow and of Hydrodynamic
Interactions}
\date{\today}
\maketitle

\begin{abstract}
We consider a simple model for steady-state luminescence of single polymer 
chains in a dilute solution in the case when excitation quenching is due 
to energy transfer between  a donor and an acceptor attached to the ends of 
the chain. We present numerical results for Rouse chains without or with 
hydrodynamic
interactions, which are taken into account in a perturbative manner.
We consider the situations of a quiescent solvent as well as the chain  
in a shear flow and discuss the dependence of the steady-state luminescence
intensity on the strength of hydrodynamic interaction and on the shear rate
in the flow. 
\end{abstract}

\section{Introduction}

Luminescent energy transfer in polymers is an important phenomenon.
Luminescent markers are used both for probing the intrinsic polymer
dynamics, and for probing the properties of the environment
 using polymers. However, the theory of such
dynamical phenomena is to no extent satisfactory. The problem here is the
complicated nonmarkovian dynamics of the system, where the most interesting
phenomena take place on the time-scales on which the systems shows strong
memory effects. Even the corresponding initial condition problem is hard to
solve. No satisfactory quantitative theory exists at present for the
stationary case.

Let us start from formulating the problem, and discuss the simplest energy
transfer model, which will be used throughout the article. Let us assume
that the ends of a polymer are marked by a donor and an acceptor monomers.
The molecule is under constant irradiation at a resonant probe frequency, so
that the donor can get excited with probability $\lambda $ per unit time (we
consider $\lambda $ as effective intensity of the irradiation). The
relaxation of the excited state due to spontaneous emission, as well as
nonlinear effects connected with possible multiple excitation are neglected,
so that the only mechanism of relaxation is the donor-acceptor energy
transfer. We assume that the corresponding energy transfer is accompanied by
emission of a photon with the frequency different from one of the
irradiating light. This transfer takes place when donor and acceptor
approach each other at distance $a$, hereafter called reaction radius.
Physically, two situations may take
place: Being in vicinity of the acceptor the donor still can be excited, and
immediately emits a photon at the observation frequency. Another situation
is the one when, being close to the acceptor, the donor gets out of
resonance with the probe and cannot be excited. In this case, the
donor-acceptor system may be in one of the two states, {\it on} and {\it off};
being in the {\it on}-state the system may be excited with probability 
$\lambda$ per unit time, and emits the photon under the transition into the 
{\it off}-state.
In what follows the expressions {\it on} and {\it off} will be simply
used for denoting states in which the end-to-end distance is above and below
the reaction radius $a$, respectively.

The overall situation might seem simple, however it is much more complicated
than the case of irreversible cyclization 
\cite{WF1,Doi,SSS,DeGennes,PZS,BG1,S1} and is
extremely awkward for theoretical investigation, even in the absence
of flow. Our knowledge about the reaction kinetics under flow is sporadic
even for simpler reactions, see \cite{FL,KMF}.

The whole problem would be easily solvable if the
life time distributions in the {\it on}- and the {\it off}-states were known.
Then the probability to be excited being in the {\it on}-state and therefore
the intensity of the emitted light could be easily calculated. The
probabilities to be in the either state are connected with the
level-crossing properties of the random process $r(t)$, where $r(t)$ is the
instantaneous end-to-end distance of the polymer. As for all diffusive
processes, however, the level-crossing process by $r(t)$ shows a fractal
structure, so that the mean time between two such crossings is zero (this
follows immediately from the Rice formula for level crossing density and
from the form of the two-time correlation function of the end-to-end
distances, say, for a Rouse polymer, which function lacks the second
derivative at zero). Again, as for all diffusive processes, this leads to a
''tremor'' in which $r(t)$ crosses the $a$-level may times until it leaves
and performs a long excursion to either side. This ''tremor'' is due to the
fact that the diffusion approximation (Wiener process) used in the
description of the chain (for example through the Rouse-like Langevin dynamics)
does not adequately mirror a short-time dynamics of whatever physical system
\cite{Eisenberg}. However, the existence of this theoretical problem does to no
extent require for the change of the model (say, by introducing underdamped
dynamics, as proposed in \cite{Eisenberg}) since the physical problem at hand
does not depend on the too-small time behavior of the $r(t)$-process.
Indeed, this process is randomly sampled at times $t_{i}$ given by a
Poissonian flow of photons following with the rate $\lambda $. The behavior
of $r(t)$ at times much smaller than $\lambda ^{-1}$ thus cannot be sampled
and can physically play no role: this statement is a close analogue of the
Nyquist's sampling theorem. Thus, the absence of the
life-time distributions is not a problem of our theoretical model, but
a problem of standard mathematical tools which rely too much on unphysical,
but absolutely unimportant short-time properties of a Wiener process.

Therefore in what follows we concentrate on the numerical investigation of
the proposed model, and consider the intensity of stationary luminescence of the
polymer $I(\lambda )$ under constant irradiation. We discuss the Rouse model
without hydrodynamic interactions, as well as the role of hydrodynamic
interaction between the monomers, and consider the case when the polymer
molecule undergoes deformation in a (weak enough) shear flow, which does not
however cause the full stretching of the molecule. This situation is
especially interesting as the case when the stationary luminescence of
diluted polymer solution can be used as a probe for the flow structure.
Authors are not aware of any experimental realizations of such visualization
method, thus our theoretical study might serve as a proof-of-principle for
such immediate flow diagnostics method.

\section{Simulation approach}

Let us start from discussing our numerical algorithm. Our simulations
consist of two independent parts: the simulation of the $r(t)$-trajectories,
which are then stored with high enough resolution, and their analysis giving
the steady-state luminescence intensity. The reason for this approach is
that one realization of the process can then be used for getting $I(\lambda)$
for a variety of parameters $a$ and $\lambda $ of the model, so that the
most time-consuming part of the simulation has to be done only once for
exactly the time necessary to get enough statistics.

Let us concentrate first of the last part of the problem, namely on the
evaluation of the stationary luminescence intensity for a given realization
of $r(t)$-process. From the record of the $r(t)$ (time resolution of stored
data has to be much smaller than the minimal $\lambda ^{-1}$ used in
simulations) we define the $a$-crossings of the process and, for given $a$,
obtain the lengths of {\it on}- and {\it off}-intervals, which are ordered and
stored. According to the Poisson statistics, the probability not to get
excited during the {\it on}-interval of duration $t_{on}$ is exactly $\exp
(-\lambda t_{on})$, thus the probability to emit light after the
{\it on}-excursion is equal to $1-\exp (-\lambda t_{on})$. Since the intensity of
emitted light is proportional to the overall number of the intervals during
which the system made a transition into its excited state, we have for the
model where the {\it off}-state is not excitable: 
\begin{equation}
I(\lambda )=\frac{1}{T}\sum_{i=1}^{n(a,T)}\left[ 1-\exp (-\lambda
t_{i})\right]   \label{eq1}
\end{equation}
where $i$ numbers the {\it on}-intervals, $n(a,T)$ is their overall number,
which depends on the reaction radius $a$ and on the overall time of
simulations $T$. Equation (\ref{eq1}) shows that the intervals of very
small duration are sampled with the probability proportional to their
lengths so that, as anticipated, the fractal structures in vicinity of the concentration points of
the level-crossings are not resolved and play no role. Using Eq.(\ref{eq1}) it is
possible to scan the whole range of intensities $\lambda $ within one run,
which is necessary to detect nonlinear effects. The situation in which,
being in the {\it off}-state, the molecule immediately emits light,
can be taken into account by adding the corresponding intensity to 
the expression given by Eq.(\ref{eq1}),
\begin{equation}
I_{1}(\lambda )=I(\lambda )+\lambda P_{off}
\label{i1}
\end{equation}
where $P_{off}$ is the probability to be in the {\it off}-state, i.e. the
overall relative time spent below $a$. 
For example, for the case without flow, it is simply a
function of relative reaction radius $\rho =a/\sqrt{\left\langle
L^{2}\right\rangle }$, where $\left\langle L^{2}\right\rangle $ is the mean
end-to-end squared distance for the chain,
\begin{equation}
P_{off}(\rho )=\mbox{erf}\left( \sqrt{\frac{3}{2}}\rho \right) -\rho \sqrt{%
\frac{6}{\pi }}\exp \left( -\frac{3}{2}\rho ^{2}\right).
\label{poff}
\end{equation}
Since this simply corresponds to adding a linear function of $\lambda $ to
the results for the \textit{on-off} model, we concentrate in what follows
only on these results, given by Eq.(\ref{eq1}). This result holds for all 
situations without flow. In the situation with flow and with hydrodynamic 
interactions, it is hard to get the analytical expression 
for $P_{off}$. The numerical results following from our simulations are
presented in Tables 3 and 4. 

Let us now turn to simulation of the trajectories.

\subsection{The Rouse model}

We start from the Rouse chain as the simplest model for a polymer 
\cite{rouse,rouse2}. A Rouse
chain is a set of $N$ beads; each one, except for the two end beads, is
connected to two neighbors by a harmonic potential, so that the overall
potential energy of the system reads 
\begin{equation}
V=\sum_{i=1}^{N-1}\frac{1}{2}k|\vec{r}_{i}-\vec{r}_{i+1}|^{2},  \label{pot}
\end{equation}
where $k$ is the harmonic spring constant and $\vec{r}_{i}$ corresponds to
the position of the $i$-th bead. The end beads are connected only to one
neighbor. The equation of motion of the chain corresponds to overdamped
motion under the influence of thermal fluctuations: 
\begin{equation}
\dot{\vec{r_{i}}}=-\frac{1}{\gamma }\frac{\partial V}{\partial \vec{r}_{i}}+%
\frac{1}{\gamma }\vec{\eta}_{i},  \label{beaddyn}
\end{equation}
where $\gamma $ is the friction parameter and $\vec{\eta}_{i}$ is a
zero-mean white noise obeying the fluctuation-dissipation relation, 
\begin{equation}
\left\langle \eta _{i}^{\alpha }(t)\eta _{j}^{\beta }(t^{\prime
})\right\rangle =2k_{B}T\gamma \delta _{ij}\delta _{\alpha \beta }\delta
(t-t^{\prime }).  \label{fdr}
\end{equation}
In thermal equilibrium, the following relations following immediately from
the canonical distribution have to hold independently on the model (and are
always checked numerically as a proof of the quality of the simulation): 
\begin{eqnarray}
\left\langle E_{tot}\right\rangle  &=&\frac{3}{2}(N-1)k_{B}T \\
\left\langle d^{2}\right\rangle  &=&\sqrt{\frac{3k_{B}T}{k}} \\
\left\langle L^{2}\right\rangle  &=&\frac{3(N-1)k_{B}T}{k},
\end{eqnarray}
where $E_{tot}$ is the total energy, and $d$ and $L$ stand for the bead-to-bead 
and end-to-end distances, respectively.

We also now apply a shear flow to the system, $\vec{v}=(\alpha y,0,0)$. The
shear flow is implemented in Eqs.(\ref{beaddyn}) by including a term $%
+\alpha y_{i}$ for the motion in the $x$-coordinate of each bead \textit{i}, 
\begin{equation}
\dot{\vec{r_{i}}}=-\frac{1}{\gamma }\frac{\partial V}{\partial \vec{r}_{i}}+%
\frac{1}{\gamma }\vec{\eta}_{i}+\vec{(\alpha y_{i},0,0)}.
\label{beaddynshear}
\end{equation}
The characteristic intensity of the flow necessary to compare its effects on 
the chain's conformation in different situations is given by the value of 
the dimensionless parameter $\alpha \tau_R$ with $\tau_R$ being the Rouse time \cite{S1}.

\subsection{Hydrodynamic interactions}

The situation under hydrodynamic interactions is much more involved. The
standard approaches \cite{fix,rze,rze2} are very accurate but slow, 
so that we prefer an
approximate perturbative one. The quality of the corresponding
approximations is checked by calculating two thermodynamically fixed
parameters of the chain in quiescent solvent: its mean end-to-end distance
and the overall energy. We anticipate that especially the end-to-end distance in the
chain was found to be extremely sensitive to improper incorporation of the
hydrodynamic interaction. We confined ourselves to the situations under
which the first order of the perturbation theory was found sufficient.

The hydrodynamic interactions among the beads are modeled within the Zimm
scheme \cite{zimm}. 
The Zimm model is based on the Rouse chain model but the equations
of motion for different beads are coupled to each other not only through 
elastic forces but also through hydrodynamic forces. Such coupling 
is a long-range one and is introduced
through the Oseen tensor \cite{oseen}, that is a $3\times 3$ tensor defined for each pair
of beads (\textit{i}-\textit{j}), 
\begin{eqnarray}
{\mathcal{H}}_{ij} &=&\frac{1}{8\pi \eta |\vec{r}_{ij}|}\left[ \vec{%
r^{\prime }}_{ij}\left( \vec{r^{\prime }}_{ij}\right) ^{T}+{\mathcal{I}}%
\right]  \\
{\mathcal{H}}_{ii} &=&\frac{1}{\gamma }{\mathcal{I}},
\end{eqnarray}
where $\mathcal{I}$ is a unit matrix, $\vec{r}^{\prime }_{ij}$ is a
unit vector $\vec{r}_{ij}/|\vec{r}_{ij}|$ in the direction of $\vec{r}_{ij}$
and $\left( \vec{r^{\prime }}_{ij}\right) ^{T}$ is its transpose. The
viscosity parameter $\eta $ can be expressed through $\gamma $ and the
bead's size $r_{0}$ since for $i=j$ one has $1/6\pi \eta r_{0}=1/\gamma $.
Then, 
\begin{equation}
{\mathcal{H}}_{ij}=\frac{3r_{0}}{4\gamma |\vec{r}_{ij}|}\left[ \vec{%
r^{\prime }}_{ij}\left( \vec{r^{\prime }}_{ij}\right) ^{T}+{\mathcal{I}}%
\right] .
\end{equation}
In what follows we use $\gamma =1$. The equation of motion for the $i$-th
bead thus reads:

\begin{equation}
\dot{\vec{r}_{i}}=\sum_{j=1}^{N}{\mathcal{H}}_{ij}\left( \frac{\partial V}{%
\partial \vec{r}_{j}}+\vec{\eta}_{j}\right) .  \label{beaddynoseen}
\end{equation}
The noises $\vec{\eta}_{j}$
acting on different beads are now not independent, otherwise the
fluctuation-dissipation theorem would be violated. One often writes the
corresponding equation of motion in the form 
\begin{equation}
\dot{\vec{r}_{i}}={\mathcal{H}}\vec{f}_{i}+2k_{B}T{\mathcal{A}}\vec{\psi},
\end{equation}
where ${\mathcal{H}}$ is the $3N\times 3N$ matrix with the diagonal elements
being unity (in the units where $\gamma =1$) and with the nondiagonal
elements denoting the Oseen terms between the corresponding components of
velocity of different beads, and the matrix ${\mathcal{A}}=\sqrt{{\mathcal{H}%
}}$ is defined through ${\mathcal{A}}\cdot {\mathcal{A}}^{T}={\mathcal{H}}$.
The elements of the vector $\vec{\psi}$ are now independent, zero mean
Gaussian white noises. Actually, the computation of the equations of motions
in the Euler scheme reads, 
\begin{equation}
\left( 
\begin{tabular}{c}
$x_{1}(t+\Delta t)$ \\ 
$y_{1}(t+\Delta t)$ \\ 
$z_{1}(t+\Delta t)$ \\ 
$x_{2}(t+\Delta t)$ \\ 
$\cdots $ \\ 
$\cdots $ \\ 
$\cdots $%
\end{tabular}
\right) =\left( 
\begin{tabular}{c}
$x_{1}(t)$ \\ 
$y_{1}(t)$ \\ 
$z_{1}(t)$ \\ 
$x_{2}(t)$ \\ 
$\cdots $ \\ 
$\cdots $ \\ 
$\cdots $%
\end{tabular}
\right) +\Delta t{\mathcal{H}}\left( 
\begin{tabular}{c}
$f_{1}^{x}(t)$ \\ 
$f_{1}^{y}(t)$ \\ 
$f_{1}^{z}(t)$ \\ 
$f_{2}^{x}(t)$ \\ 
$\cdots $ \\ 
$\cdots $ \\ 
$\cdots $%
\end{tabular}
\right) +\sqrt{2k_{B}T\Delta t}{\mathcal{A}}\left( 
\begin{tabular}{c}
$\psi _{1}^{x}(t)$ \\ 
$\psi _{1}^{y}(t)$ \\ 
$\psi _{1}^{z}(t)$ \\ 
$\psi _{2}^{x}(t)$ \\ 
$\cdots $ \\ 
$\cdots $ \\ 
$\cdots $%
\end{tabular}
\right) ,
\label{eqsdyn}
\end{equation}
where $f_{i}^{\beta }$ are the forces due to the harmonic springs for the 
\textit{i}-th bead in the $\beta $ axis and $\psi _{i}^{\beta }$ are the
corresponding components of $\vec{\psi}$.

The computation of ${\mathcal{A}}$ can be performed exactly by diagonalizing 
${\mathcal{H}}$. This exact diagonalization requires an extremely high computational cost
for long chains. The widely used method based on the orthogonal polynomials
decomposition (which gives very exact results) is still too slow to get the
runs long enough for our purposes. Therefore we decided for a simple
approximate approach based on the perturbation expansion of the hydrodynamic
interaction.

To do this we write ${\mathcal{H}}$ as ${\mathcal{I}}+r_{0}{\mathcal{S}}$,
and then expand the square root ${\mathcal{A}}=\sqrt{{\mathcal{I}}+r_{0}{%
\mathcal{S}}}$ in powers of $r_{0}$, 
\begin{equation}
{\mathcal{A}}\approx {\mathcal{I}}+\frac{r_{0}{\mathcal{S}}}{2}-\frac{%
r_{0}^{2}{\mathcal{S}}^{2}}{8}+\cdots .
\end{equation}

Since in the thermal equilibrium the averages $\left\langle
E_{tot}\right\rangle $ being the internal energy and $\left\langle
L^{2}\right\rangle $ (also being a thermodynamical quantity following
immediately from equipartition) are not modified by the dissipative coupling
introduced by the Oseen tensor, we can numerically check the validity of the
approximations for ${\mathcal{A}}$ for different $r_{0}$ values. We see that 
$\left\langle L^{2}\right\rangle $ is extremely sensitive to incorrect
incorporation of the hydrodynamic interaction, and its calculation is used
as a probe of the quality of the approximation, see Tables 1 and 2. The data
for Rouse model give us typical error bars for the simulation of the exact
model on the same scale.

Looking at the Tables 1 and 2, one can conclude that for $r_{0}$ up to $0.2$, the second
order approximation is sufficient, and for $r_{0}$ up to $0.1$, the first
order approximation (much shorter simulations) is accurate enough. In the case
$r_{0} = 0.5$ also the second order gets insufficient. Thus, in our simulations
we restrict ourselves to $r_{0} \leq 0.2$.  
We use a second order Runge-Kutta method to solve Eqs.(\ref{eqsdyn}) with a
sufficiently small time step $\Delta t =10^{-3}$. For the results shown
in this paper we run $2 \cdot 10^{7}$ iterations up to a maximum time 
$t = 2 \cdot 10^{4}$
for a full trajectory needed for adequate statistics. An initial 
thermalization period of $1000$ time units is performed in all cases 
in order to start the trajectories from a thermal equilibrium state.
The {\it Compaq AlphaServer HPC320} used to run these simulations requires 
about $3$ hours of CPU time
for $N=51$ when the first order approximation scheme is chosen. The second
perturbative order requires more than $120$ hours
of CPU time for the same number of iterations and chain length.

\begin{table}{{\bf Table 1:} The quality of perturbative approximations for 
$N=21$}
\begin{center}
\begin{tabular}{|c|c|c|c|}
\hline
 &  & $<E_{tot}>$ & $<L^{2}>$ \\ \hline
Theoretical &  & 30 & 20 \\ \hline
Rouse simulation &  & 30.185 & 21.028 \\ 

 \hline
Zimm, $r_{0}=0.1$ & 
\begin{tabular}{c}
$0$-order \\ 
$1$-order \\ 
$2$-order
\end{tabular}
& 
\begin{tabular}{c}
29.591 \\ 
30.763 \\ 
30.221
\end{tabular}
& 
\begin{tabular}{c}
27.406 \\ 
21.020 \\ 
20.181
\end{tabular}
\\ \hline
Zimm, $r_{0}=0.2$ & 
\begin{tabular}{c}
$0$-order \\ 
$1$-order \\ 
$2$-order
\end{tabular}
& 
\begin{tabular}{c}
31.190 \\ 
33.114 \\ 
31.676
\end{tabular}
& 
\begin{tabular}{c}
36.923 \\ 
23.843 \\ 
21.839
\end{tabular}
\\ \hline
Zimm, $r_{0}=0.5$ & 
\begin{tabular}{c}
$0$-order \\ 
$1$-order \\ 
$2$-order
\end{tabular}
& 
\begin{tabular}{c}
42.058 \\ 
49.344 \\ 
53.215
\end{tabular}
& 
\begin{tabular}{c}
73.349 \\ 
44.741 \\ 
46.732
\end{tabular}
\\ \hline
\end{tabular}
\end{center}
\end{table}

\begin{table}{{\bf Table 2:} The quality of perturbative approximations for $N=51$} 
\begin{center}
\begin{tabular}{|c|c|c|c|}
\hline
$N=51$ &  & $<E_{tot}>$ & $<L^{2}>$ \\ \hline
Theoretical &  & 75 & 50 \\ \hline
Rouse simulation &  & 75.323 & 48.122 \\ \hline
Zimm, $r_{0}=0.1$ & 
\begin{tabular}{c}
$0$-order \\ 
$1$-order \\ 
$2$-order
\end{tabular}
& 
\begin{tabular}{c}
76.063 \\ 
77.412 \\ 
75.926
\end{tabular}
& 
\begin{tabular}{c}
86.674 \\ 
54.021 \\ 
47.746
\end{tabular}
\\ \hline
Zimm, $r_{0}=0.2$ & 
\begin{tabular}{c}
$0$-order \\ 
$1$-order \\ 
$2$-order
\end{tabular}
& 
\begin{tabular}{c}
81.399 \\ 
83.903 \\ 
79.167
\end{tabular}
& 
\begin{tabular}{c}
130.64 \\ 
67.487 \\ 
51.600
\end{tabular}
\\ \hline
\end{tabular}
\end{center}
\end{table}

\section{Results}

Although the overall role of flow and hydrodynamical interaction is rather clear,
the behavior of the intensity as a function of parameters $\alpha$ and $r_0$
is not trivial. The flow elongates the molecule, so that the typical end-to-end 
distance grows with $\alpha$ while the hydrodynamical interactions slow-down
the dynamics of intramolecular relative motion, which increases the characteristic 
time spent in {\it on}-state.  

\begin{figure}[tbh]
\begin{center}
\epsfxsize = 3.5in
\epsffile{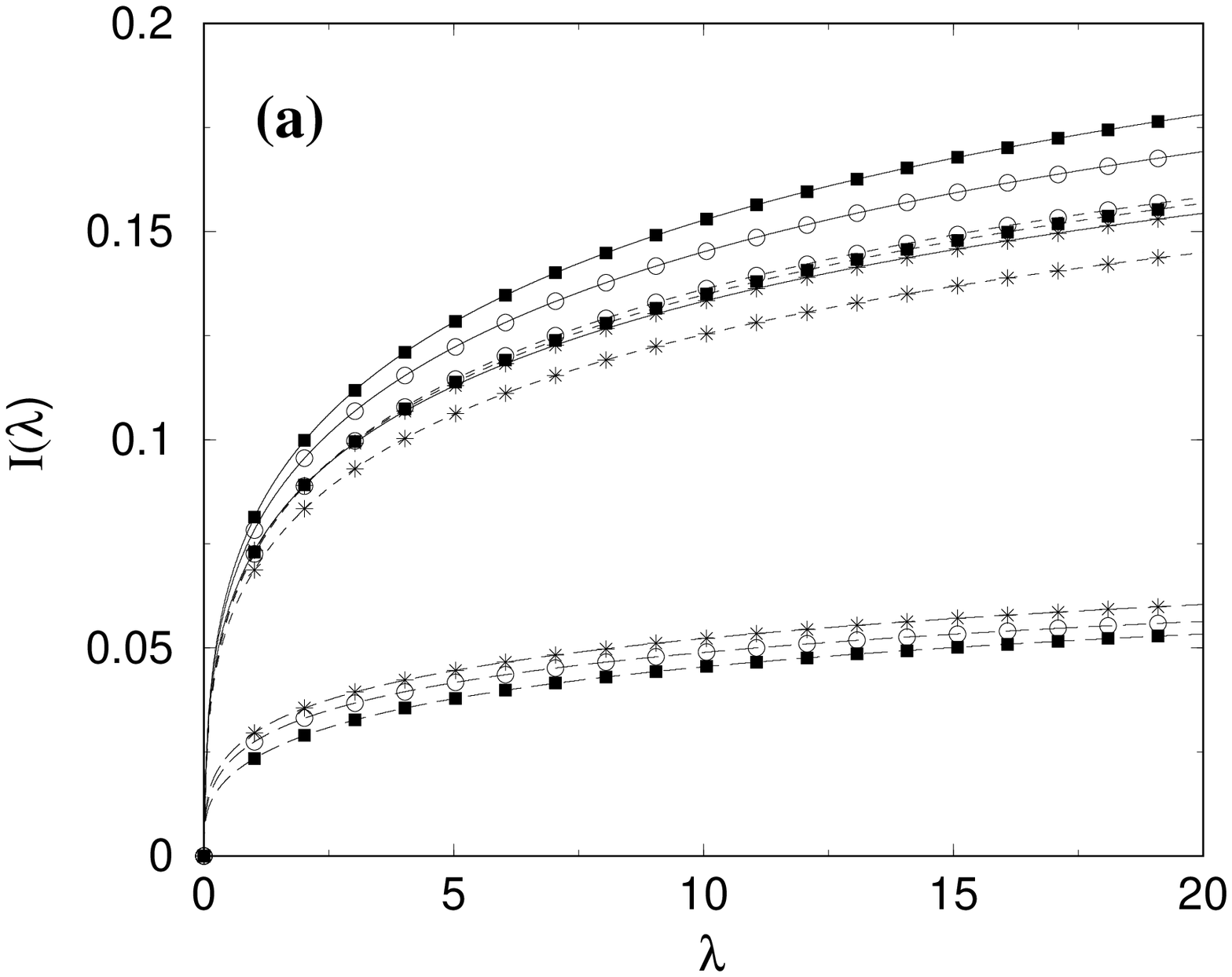}
\epsfxsize = 3.5in 
\epsffile{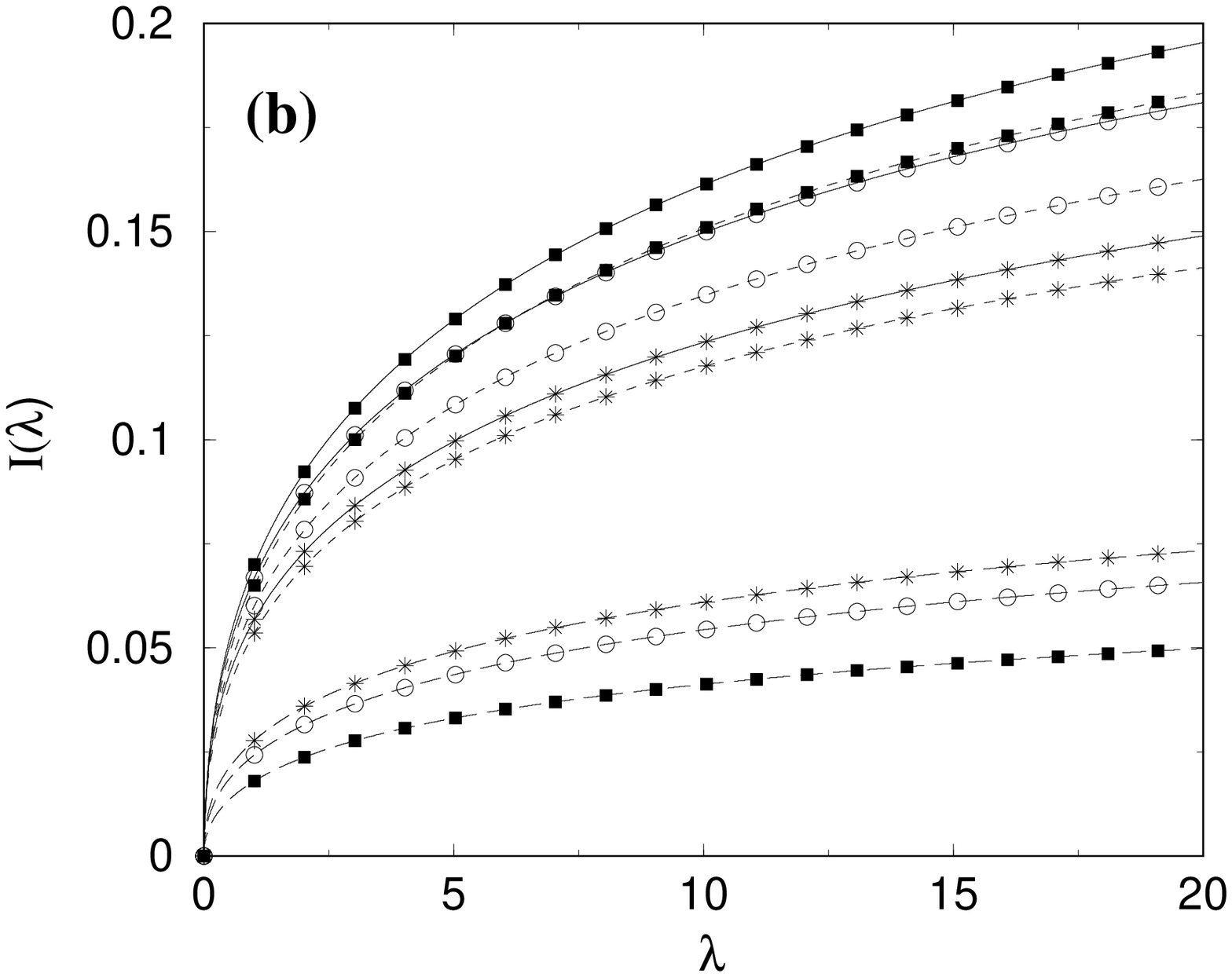}
\end{center}
\caption{The intensity of steady-state luminescence in the {\it on-off} model
as a function of irradiation intensity $\lambda$ for a reaction radius such that $\rho \ll 1$.
In panel (a) $N=21$ and $a=1$. In panel (b) $N=51$ and $a=2$.
In both panels the same notation is used:
the symbol indicates the value of the bead radius $r_0$:
filled squares ($0$), empty circles ($0.05$) and starts ($0.1$),
whereas solid, dotted and dashed lines correspond to
$\alpha \tau_{R}=0$, $1.05$ and $6.56$, respectively.
$\tau_R=13.51$ for the chain with $21$ beads and $\tau_R=84.43$
for chains with $51$ beads.}
\label{figsmallro}
\end{figure}

The behavior of $I(\lambda)$ as a function of hydrodynamic radius and flow
intensity strongly depends on the relation $\rho$ between the reaction radius, $a$,
and the equilibrium end-to-end distance of the polymer
(i.e. the one in the absence of the flow), $L=\sqrt{\left\langle L^{2}\right\rangle}$.
For $\rho \ll 1$ the polymer
is typically in the \textit{on}-state, and thus the flow (elongating the chain and
making the transition into the \textit{off}-state less probable) and
the hydrodynamic interaction without flow (making the change of states slower) 
work in the same direction and lead to the decrease in intensity, as it is 
clearly seen in Fig.\ref{figsmallro}. 

\begin{figure}[tbh]
\begin{center}
\epsfxsize = 3.5in 
\epsffile{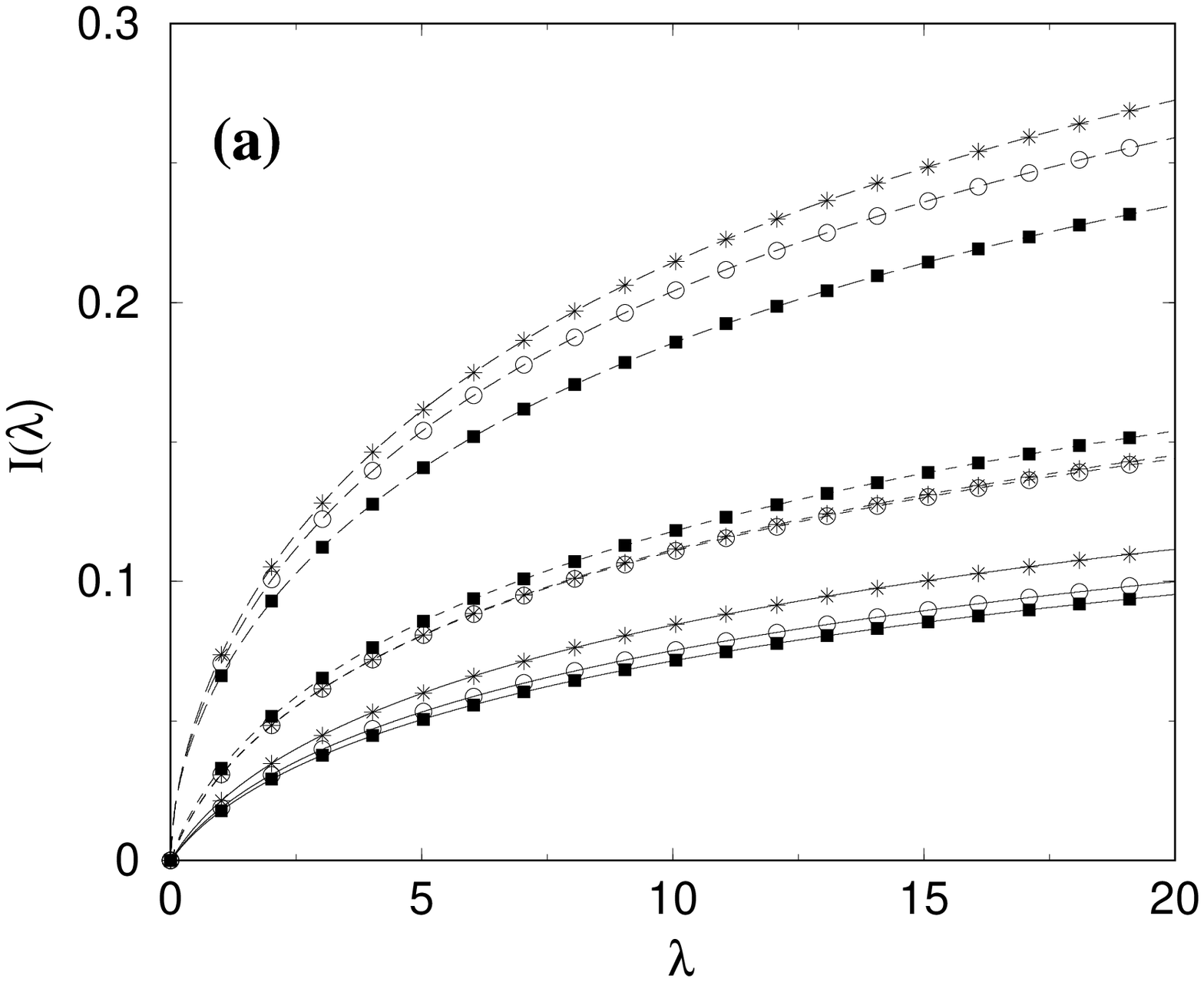}
\epsfxsize = 3.5in 
\epsffile{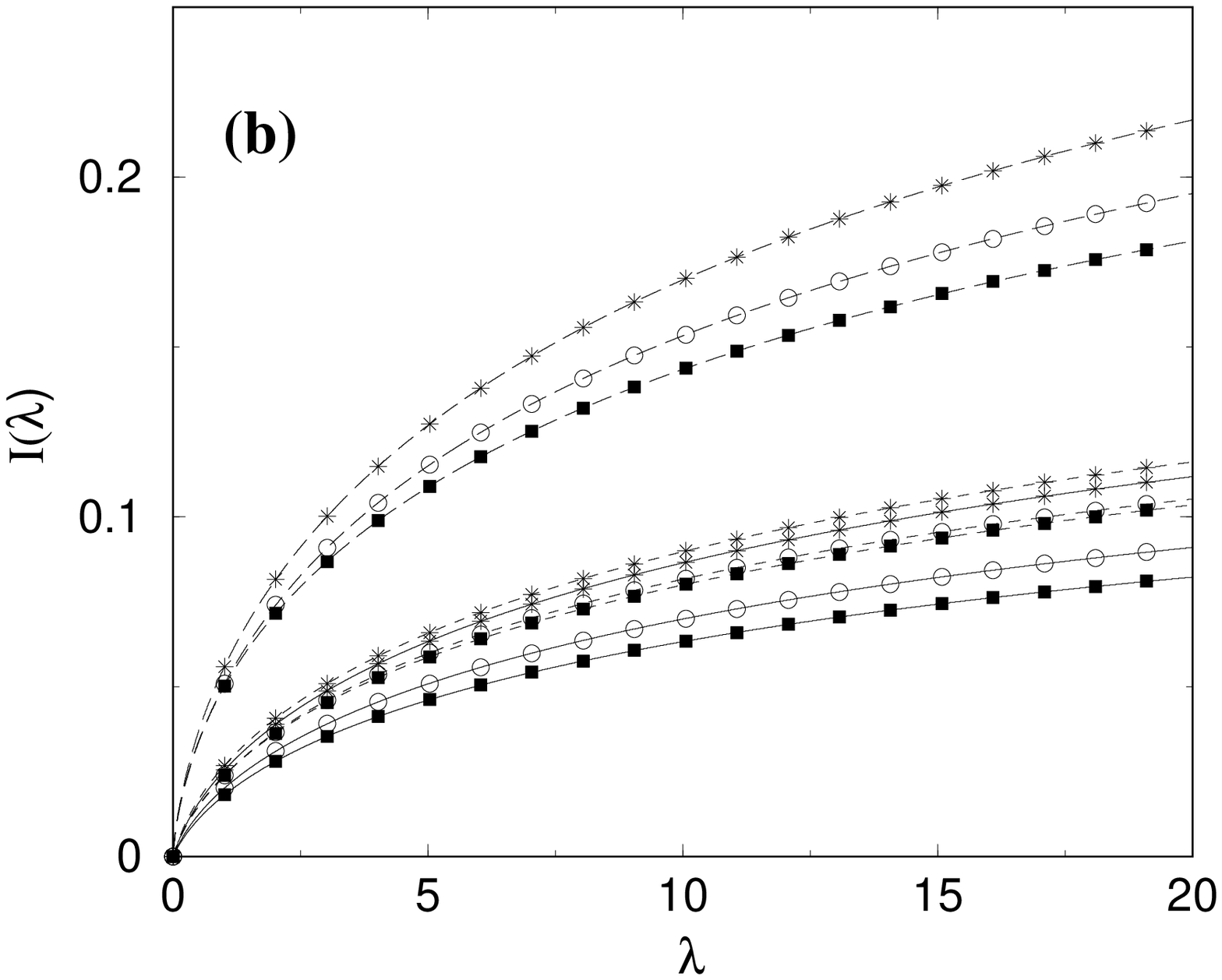}
\end{center}
\caption{Same as in Fig.\ref{figsmallro}, but now for $\rho \gg 1$.
In panel (a) $N=21$ and $a=8$. In panel (b) $N=51$ and $a=12$.
We use the same notation for the lines as in Fig.\ref{figsmallro}}
\label{figlargero}
\end{figure}

For $\rho \gg 1$ the molecule is typically in the \textit{off}-state. Increasing flow
increases the probability of switching to the the \textit{on}-state, and thus 
leads to increase in the steady-state intensity. The hydrodynamic interaction 
in the absence of the flow also leads to increasing the typical time in the corresponding
state. The effects of the flow and
the hydrodynamic interactions for $\rho \gg 1$ are depicted in Fig.\ref{figlargero}. 
The increasing effect of the hydrodynamic interaction has to
do with the interplay of two factors. On the one hand, the longer \textit{on}-intervals
get even longer under hydrodynamic interaction, and thus give smaller contributions
to the the overall intensity. On the other hand, increasing the interaction makes that
more shorter {\it on}-intervals are now resolved on the timescale 
of $\lambda^{-1}$, and these contributions in the intensity overweigh the loss
due to the former of both effects.

This explanation shows that the role of hydrodynamic interaction is rather
subtle, and may lead to interesting effects for both regimes
($\rho \ll 1$ and $\rho \gg 1$), especially when the flow is present.  
Indeed, the effect of hydrodynamic interaction for the cases with $\alpha \neq 0$
depends in a fine way on all parameters, and may act in opposite directions
(compare the curves for no flow and high flow, $\alpha \tau_R = 6.56$, in both
panels of Fig.\ref{figsmallro}, and
the curves for no flow and moderate flow, $\alpha \tau_R = 1.05$,
in the upper panel of Fig.\ref{figlargero}). 

The values of $P_{off}$ which are necessary to establish the connection 
between the two situations discussed in the Introduction (Eq.(\ref{i1})) are given in
Tables 3 and 4 for $N=21$ and for $N=51$, respectively. 
The intensities of the flows in these tables correspond to the same values of the dimensionless
flow intensities $\alpha \tau_R=0, 0.176, 1.05$ and $6.56$ for $N=21$ and 
for  $N=51$ chains (the Rouse times being $\tau_R=13.51$ and $\tau_R=84.43$, respectively).

\section{Conclusions}

We presented the results of numerical simulations of the intensity of
steady-state luminescence of single polymer chains in a dilute solution
due to excitation quenching in a simple model in which donor and acceptor 
are attached to the ends of the chain.
The chain is modeled by simple Rouse dynamics without or with hydrodynamic
interactions, which are taken into account in a perturbative manner.
We consider the situations of a quiescent solvent as well as the chain  
in a shear flow. Depending on the relation between 
the effective distance for energy transfer and the typical end-to-end distance 
of the chain different regimes are encountered with respect to dependence
of the steady-state luminescence intensity on the strengths of the flow
and of interaction. Such luminescent probes may be used for experimental
flow diagnostics.

\section{Acknowledgments}

The authors acknowledge helpful discussions with J.M. Sancho and F. Sagues. 
We thank CESCA (Centre de Supercomputaci\'{o} de Catalunya) for financial and
computational support through the 'Improving the Human Potential' Program.
IMS gratefully acknowledges partial financial support by the Fonds der Chemischen Industrie.

\begin{table}{{\bf Table 3:} $P_{off}$ for the chain with $N=21$} \\[1ex]
\begin{center}
\begin{tabular}{|c|c|c|c|} \hline
 $\alpha =0$ & $a=1$ &  $a=4$ &   $a=8$    \\
\hline
\begin{tabular}{c} $r_0=0$ \\  $r_0=0.05$  \\  $r_0=0.1$  \end{tabular} &
\begin{tabular}{c} 0.014843 \\  0.014831  \\  0.013845  \end{tabular} & 
\begin{tabular}{c} 0.502626 \\  0.499207  \\  0.484438  \end{tabular} & 
\begin{tabular}{c} 0.974239 \\  0.972863  \\  0.969214  \end{tabular}  \\
\hline
 $\alpha =0.0125$ & $a=1$ &  $a=4$ &   $a=8$    \\
\hline
\begin{tabular}{c} $r_0=0$ \\  $r_0=0.05$  \\  $r_0=0.1$  \end{tabular} &
\begin{tabular}{c} 0.01467 \\  0.01481  \\  0.013724  \end{tabular} & 
\begin{tabular}{c} 0.49977 \\  0.49775  \\  0.48283  \end{tabular} & 
\begin{tabular}{c} 0.97302 \\  0.97191  \\  0.96835  \end{tabular}  \\
\hline
 $\alpha =0.078$ & $a=1$ &  $a=4$ &   $a=8$    \\
\hline
\begin{tabular}{c} $r_0=0$ \\  $r_0=0.05$  \\  $r_0=0.1$  \end{tabular} &
\begin{tabular}{c} 0.013605 \\  0.013742  \\  0.012997  \end{tabular} & 
\begin{tabular}{c} 0.443851 \\  0.456638  \\  0.450992  \end{tabular} & 
\begin{tabular}{c} 0.937554 \\  0.947050  \\  0.949065  \end{tabular}  \\
\hline
 $\alpha =0.488$ & $a=1$ &  $a=4$ &   $a=8$    \\
\hline
\begin{tabular}{c} $r_0=0$ \\  $r_0=0.05$  \\  $r_0=0.1$  \end{tabular} &
\begin{tabular}{c} 0.006830 \\  0.007423  \\  0.008600  \end{tabular} & 
\begin{tabular}{c} 0.15618 \\  0.18520  \\  0.20918  \end{tabular} & 
\begin{tabular}{c} 0.43874 \\  0.52606  \\  0.58951  \end{tabular}  \\
\hline
\end{tabular}
\end{center}
\end{table}

\begin{table}{{\bf Table 4:} $P_{off}$ for the chain with $N=51$}\\[1ex]
\begin{center}
\begin{tabular}{|c|c|c|c|} \hline
 $\alpha =0$ & $a=2$ &  $a=7$ &   $a=12$    \\
\hline
\begin{tabular}{c} $r_0=0$ \\  $r_0=0.05$  \\  $r_0=0.1$  \end{tabular} &
\begin{tabular}{c} 0.034207 \\  0.033190  \\  0.028929  \end{tabular} & 
\begin{tabular}{c} 0.613049 \\  0.594208  \\  0.556314  \end{tabular} & 
\begin{tabular}{c} 0.960062 \\  0.957482  \\  0.945990  \end{tabular}  \\
\hline
 $\alpha =0.002$ & $a=2$ &  $a=7$ &   $a=12$    \\
\hline
\begin{tabular}{c} $r_0=0$ \\  $r_0=0.05$  \\  $r_0=0.1$  \end{tabular} &
\begin{tabular}{c} 0.03386 \\  0.03270  \\  0.02932  \end{tabular} & 
\begin{tabular}{c} 0.61418 \\  0.59412  \\  0.55641  \end{tabular} & 
\begin{tabular}{c} 0.95983 \\  0.95756  \\  0.94633  \end{tabular}  \\
\hline
 $\alpha =0.0125$ & $a=2$ &  $a=7$ &   $a=12$    \\
\hline
\begin{tabular}{c} $r_0=0$ \\  $r_0=0.05$  \\  $r_0=0.1$  \end{tabular} &
\begin{tabular}{c} 0.033417 \\  0.030730  \\  0.027587  \end{tabular} & 
\begin{tabular}{c} 0.575235 \\  0.571878  \\  0.540211  \end{tabular} & 
\begin{tabular}{c} 0.933795 \\  0.945019  \\  0.939310  \end{tabular}  \\
\hline
 $\alpha =0.078$ & $a=2$ &  $a=7$ &   $a=12$    \\
\hline
\begin{tabular}{c} $r_0=0$ \\  $r_0=0.05$  \\  $r_0=0.1$  \end{tabular} &
\begin{tabular}{c} 0.01609 \\  0.02417  \\  0.01273  \end{tabular} & 
\begin{tabular}{c} 0.20126 \\  0.28267  \\  0.30860  \end{tabular} & 
\begin{tabular}{c} 0.45965 \\  0.61573  \\  0.67650  \end{tabular}  \\
\hline
\end{tabular}
\end{center}
\end{table}

\newpage

\end{document}